\documentclass[12pt,a4paper,final]{iopart}

\usepackage{graphicx}
\usepackage[breaklinks=true,colorlinks=true,linkcolor=blue,urlcolor=blue,citecolor=blue]{hyperref}
\usepackage{natbib}
\setcitestyle{numbers}
\usepackage{filecontents}
\usepackage{subcaption}
\usepackage{float}

\begin{document}

\title[Percolation Thresholds for Robust Network Connectivity]{Percolation Thresholds for Robust Network Connectivity}

% \begin{flushleft}
\author[]{}\textbf{Arman Mohseni-Kabir$^1$, Mihir Pant$^2$, Don Towsley$^3$, Saikat Guha$^4$, Ananthram Swami$^5$}
\begin{flushleft}
{$1$. Physics Department, University of Massachusetts Amherst, arman@physics.umass.edu} 

{$2$. Massachusetts Institute of Technology}\par
{$3$. College of Information and Computer Sciences, University of Massachusetts Amherst}\par
{$4$. College of Optical Sciences, University of Arizona}\par
{$5$. Army Research Laboratory}
\end{flushleft}

    % \makebox[0.005cm] \bf\author{Arman Mohseni-Kabir$^1$, Mihir Pant$^2$, Don Towsley$^3$, Saikat Guha$^4$, Ananthram Swami$^5$}\par
    % \makebox[0.1cm]\address{$1$. Physics Department, University of Massachusetts Amherst, arman@physics.umass.edu} \par
    % \makebox[0.1cm]\address{$2$. Massachusetts Institute of Technology}\par
    % \makebox[0.1cm]\address{$3$. College of Information and Computer Sciences, University of Massachusetts Amherst}\par
    % \makebox[0.1cm]\address{$4$. College of Optical Sciences, University of Arizona}\par
    % \makebox[0.1cm]\address{$5$. Army Research Laboratory}\par

\section*{Abstract}
Communication networks, power grids, and transportation networks are all examples of networks whose performance depends on reliable connectivity of their underlying network components even in the presence of usual network dynamics due to mobility, node or edge failures, and varying traffic loads. Percolation theory quantifies the threshold value of a local control parameter such as a node occupation (resp., deletion) probability or an edge activation (resp., removal) probability above (resp., below) which there exists a giant connected component (GCC), a connected component comprising of a number of occupied nodes and active edges whose size is proportional to the size of the network itself. Any pair of occupied nodes in the GCC is connected via at least one path comprised of active edges and occupied nodes. The mere existence of the GCC itself does not guarantee that the long-range connectivity would be {\em robust}, e.g., to random link or node failures due to network dynamics. In this paper, we explore new percolation thresholds that guarantee not only spanning network connectivity, but also robustness. We define and analyze four measures of robust network connectivity, explore their interrelationships, and numerically evaluate the respective robust percolation thresholds for the 2D square lattice.

\vspace{2pc}
\noindent{\it Keywords}: Percolation, Network Robustness, Phase Transition, Critical Phenomena
%\maketitle

\section{Introduction}\label{sec:intro}

In recent years, there has been impressive progress in our understanding of structural and dynamical properties of complex systems. Network science has emerged as a prominent field which provides us novel perspectives to better understand complexity \cite{barabasi2011network,strogatz2001exploring}. This is because many complex systems can be described with networks in which the entities are represented by nodes and the relationship between these entities are represented as bonds connecting these nodes.  The dynamics of these complex systems can be modeled through the structural dynamics of the network as well as the dynamics on the network \cite{newman2003structure,dorogovtsev2008critical,albert2002statistical}. Structural transitions in networks has been a focus of numerous research studies in the past decades due to their importance in characterizing the performance of natural and man-made networks. These structural transitions affect many  important properties of networks, e.g. robustness to breakdowns \cite{callaway2000network,cohen2000resilience}, cascading failure in networks \cite{buldyrev2010catastrophic}, and epidemic spreading on social and technological networks. \cite{vespignani2012modelling,kitsak2010identification,pastor2001epidemic}. One of the most important properties of these networks is their functional and structural robustness to unexpected interruptions caused due to node and edge failures. At the hallmark of these studies, percolation theory \cite{grimmett1999percolation} quantifies the robustness of networks by looking at the size of the largest {\em connected component} (LCC), i.e., the largest set of nodes of which each pair  is connected by at least one path, as a function of a probability or a rate parameter that controls either random \cite{cohen2000resilience}, localized \cite{shao2015percolation,berezin2015localized}, or targeted \cite{dong2012percolation,huang2011robustness} node and/or edge removals. A {\em percolation threshold} refers to the value of that parameter that separates a phase transition between two regimes of network connectedness---a set of small disconnected islands of connectivity on one side (the {\em subcritical regime}), and the existence of a giant connected component (GCC), a connected component of size proportional to the size of the network, on the other side (the {\em supercritical regime}). The sudden appearance of the GCC at this threshold is referred to as percolation. As one goes deeper in the supercritical regime, the GCC becomes progressively more richly connected. Theoretical and numerical computation of percolation thresholds has been an ongoing challenge in the scientific community \cite{stauffer2018introduction}. Due to the high complexity of the problem, analytical results exist only for very few lattice structures. On the other hand, numerical simulations has shown very effective in determining the threshold for a range of regular and disordered lattices as well as random networks. The exponential increase in computational resources has paved the way for more precise calculation of percolation thresholds \cite{wang2013bond,mertens2017percolation}.
 
In many real-world applications, {\em barely} meeting the percolation threshold may not suffice to ensure robust network operation, since the GCC may be fragile and prone to breaking with the failure of only a small fraction of nodes or edges. Further, even in the absence of actual failures, in certain networks, for example communication networks and biological networks \cite{kim2013phase}, it is often desirable not only to be long-range connected but to have multiple paths connecting pairs of nodes in order to help control network congestion and support higher data throughput. 
Therefore, there is a need to introduce more advanced measures of network robustness, which not only capture spanning connectivity but are also able to accommodate some additional measure of robustness. Clearly, to meet any such additional robustness-driven constraint, the network must be pushed deeper within the supercritical regime of standard percolation theory. Expanding on ideas from percolation theory, researchers have studied other variants of percolation on different networks including $k$-core percolation \cite{dorogovtsev2006k,yuan2016k,lee2016critical}, $k$-clique percolation \cite{li2015clique,derenyi2005clique}, and $k$-connectivity \cite{newman2008bicomponents,kim2013phase}.

In this paper, we define and perform a comparative analysis of four intuitive measures of network robustness. We explain their inter-relationships, and also numerically evaluate the respective robust percolation thresholds for the square lattice. We leave connecting our robustness measures to application-specific robustness measures in real-life networks, for future work.

\section{Robustness measures}\label{sec:definitions}

\begin{figure}
   \includegraphics[width = \columnwidth]{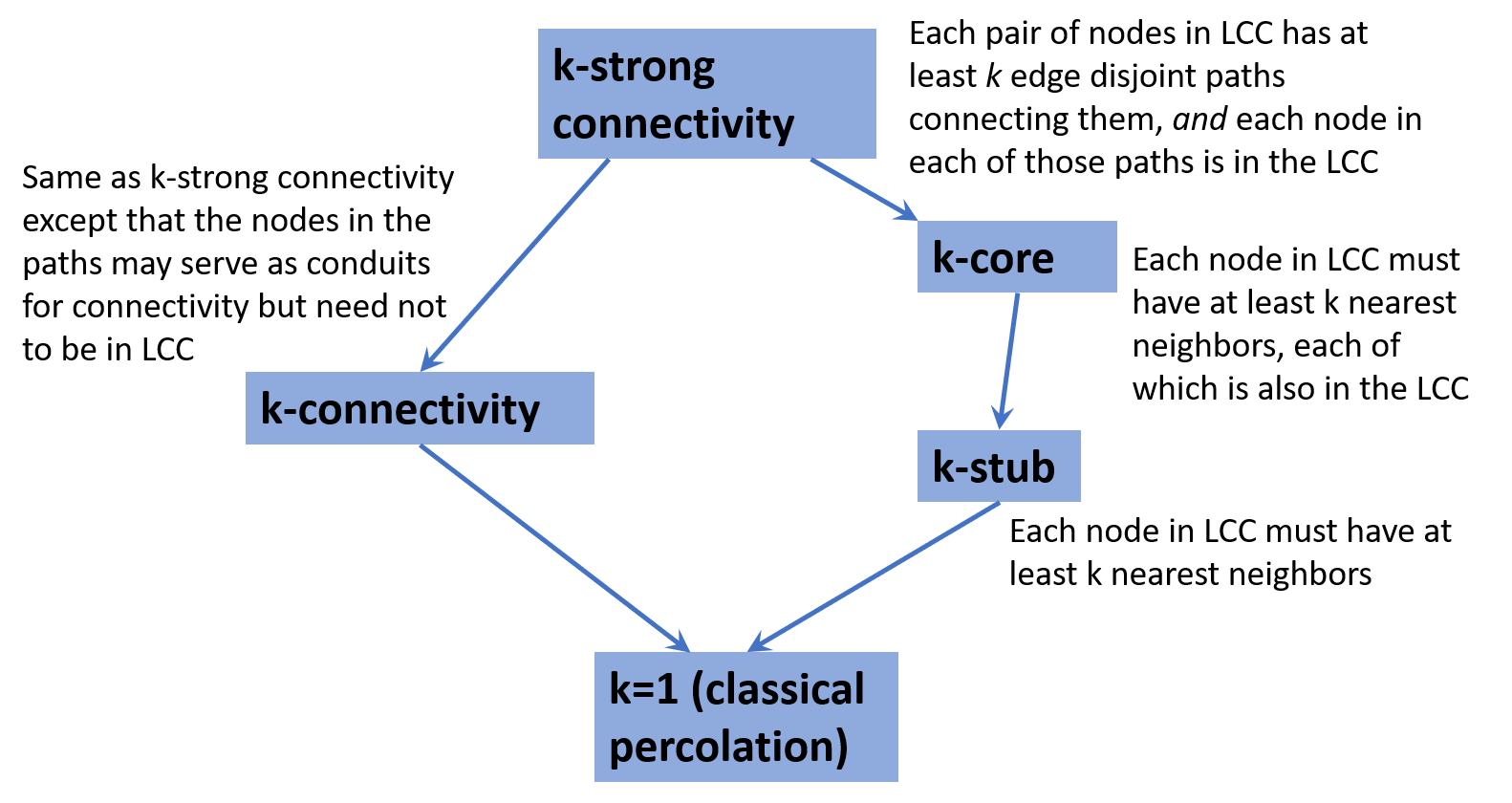}
   \caption{Models of robustness in connectivity. The arrows depict going from stronger to progressively weaker measures of robustness. In other words, if a network is in the supercritical regime with respect to the stronger of two robustness models, it will be so for the weaker one, but not necessarily vice versa.}
   \label{fig:robustness_model}
\end{figure}

The four robustness measures that we investigate in this paper are depicted in Fig.~\ref{fig:robustness_model}: (1) $k$-strong-connectivity, (2) $k$-connectivity \cite{newman2008bicomponents}, (3) $k$-core connectivity \cite{dorogovtsev2006k} and (4) $k$-stub connectivity \cite{carmi2007model}. The arrows in Fig.~\ref{fig:robustness_model} depict going from stronger to weaker measures of robustness. In other words, if a network is in the supercritical regime with respect to the stronger of two robustness models, it will also be so for the weaker one, but not necessarily vice versa. In each of the robust connectivity models, $k \in \left\{1, 2, \ldots\right\}$ denotes the strength of the robustness setting, and each model is defined in a way such that $k=1$ reduces each to the standard bond (or, respectively site) percolation model. Next to each measure of robustness in Fig.~\ref{fig:robustness_model}, we write the condition that a robust $k$-connected component must satisfy under that measure of robustness. The network is said to percolate within any given robustness measure, when the size of the respective largest robust $k$-connected component is proportional to the size of the network itself.

In this Section, we formally define and explain the intuition behind each of these robustness measures. In Section~\ref{sec:thresholds}, for each of these measures, we define bond-percolation and site-percolation thresholds with respect to each of the measures, for the 2D square lattice, and interpret our results.

\begin{enumerate}

\item {\bf $k$-strong connectivity}: This is the strongest notion of robustness that we study. In this model, a $k$-connected component is a set of nodes such that every pair has at least $k$ node-disjoint paths connecting them. The nodes in the paths must also satisfy this conditions and be a part of the k-strong component. This means that removal of $k-1$ nodes from a $k$-connected component will not disconnect the rest of the nodes in that component.

\item {\bf $k$-connectivity}:  In this model, a $k$-connected component is a set of nodes in which each pair is connected via at least $k$ node-disjoint paths. However those paths need not belong to the $k$-connected component. Newman {\em et. al}~\cite{newman2008bicomponents} showed that the percolation threshold for a configuration model random graph (any node degree distribution) is the same as that of standard percolation, although the absolute size of the GCC in the supercritical regime varies with $k$ for $k > 1$. 

\item {\bf $k$-core connectivity}: In this model, a $k$-connected component is a set of nodes such that each node has at least $k$ nearest neighbors each of which is also in the $k$-component. The concept of $k$-core connectivity and decomposing a complex network into its $k$-core components has been applied to several real-world networks, e.g., the Internet, the World Wide Web, and cellular networks~\cite{alvarez2006k}. 

\item {\bf $k$-stub connectivity}: In this model, a $k$-stub connected component is a set of nodes such that each has at least $k$ nearest neighbors (which need not belong to the $k$-component).
\end{enumerate}

For any $k > 1$, there is a hierarchical relation between these four measures of robustness---$k$-strong-connectivity being the strongest of all, and regular percolation being the weakest. Connected components under these measures of robustness are nested per the hierarchies shown in Fig.~\ref{fig:robustness_model}. For example, a $2$-core connected component under the $k$-core model is always a subgraph of a $2$-stub connected component under the $k$-stub model, which in turn is a subgraph of a regular connected component. There is no established hierarchical relationship between $k$-connectivity, and either the $k$-core or $k$-stub models.

\section{Robust percolation thresholds}\label{sec:thresholds}

\subsection{$k$-connectivity}\label{sec:kconnectivity}
$k$-connectivity is the best-known robustness measure studied in the literature. We call a subset of network nodes $k$-connected if each pair of nodes in that subset has at least $k$ edge-disjoint paths connecting them. These paths can contain nodes that act as conduits to connect two nodes in the $k$-connected component while themselves not being part of the $k$-connected component. There are several algorithms for finding $k$-connected components for different values of $k$ for a given graph. However, linear-time algorithms are only known for the cases $k = 2$ and $k = 3$ \cite{hopcroft1973dividing}. For $k>3$, polynomial-time algorithms exist to find the $k$-connected components of a graph \cite{cheriyan1991algorithms}. 
To gain a better understanding of the behavior of $k$-connected components, we restrict ourselves to site and bond percolation on the 2D square lattice. Despite percolation on 2D square lattice being a widely studied problem, to the best of our knowledge there is no literature on $k$-connectivity properties of percolation clusters (in the sub-critical and super-critical regime). Grimmett \cite{grimmett1997percolation} showed that the bond percolation threshold of the square lattice is given by $p_{c}=\frac{1}{2}$ and stated the following theorem:

Suppose $B_{n}$ is an $n$ by $n$ square grid centered at the origin. Let $M_{n}$ denote the maximal number of (pairwise) edge-disjoint left to right paths crossings $B_{n}$. Then, for any $p>p_{c}$ there exists positive constants $\eta=\eta(p)$ and $\lambda=\lambda(p)$ such that:
\begin{center}
$P_{p}(M_{n}\leq\eta n)\leq\exp(-\lambda n)$
\end{center}

Where $P_{p}(M_{n}\leq\eta n)$ denotes the probability of occurrence of event $M_{n}\leq\eta n$. This means that there exists order $n$ disjoint left-right crossings of box $B_{n}$ when $p>\frac{1}{2}$.

\begin{figure}[H]
   \includegraphics[width=\columnwidth]{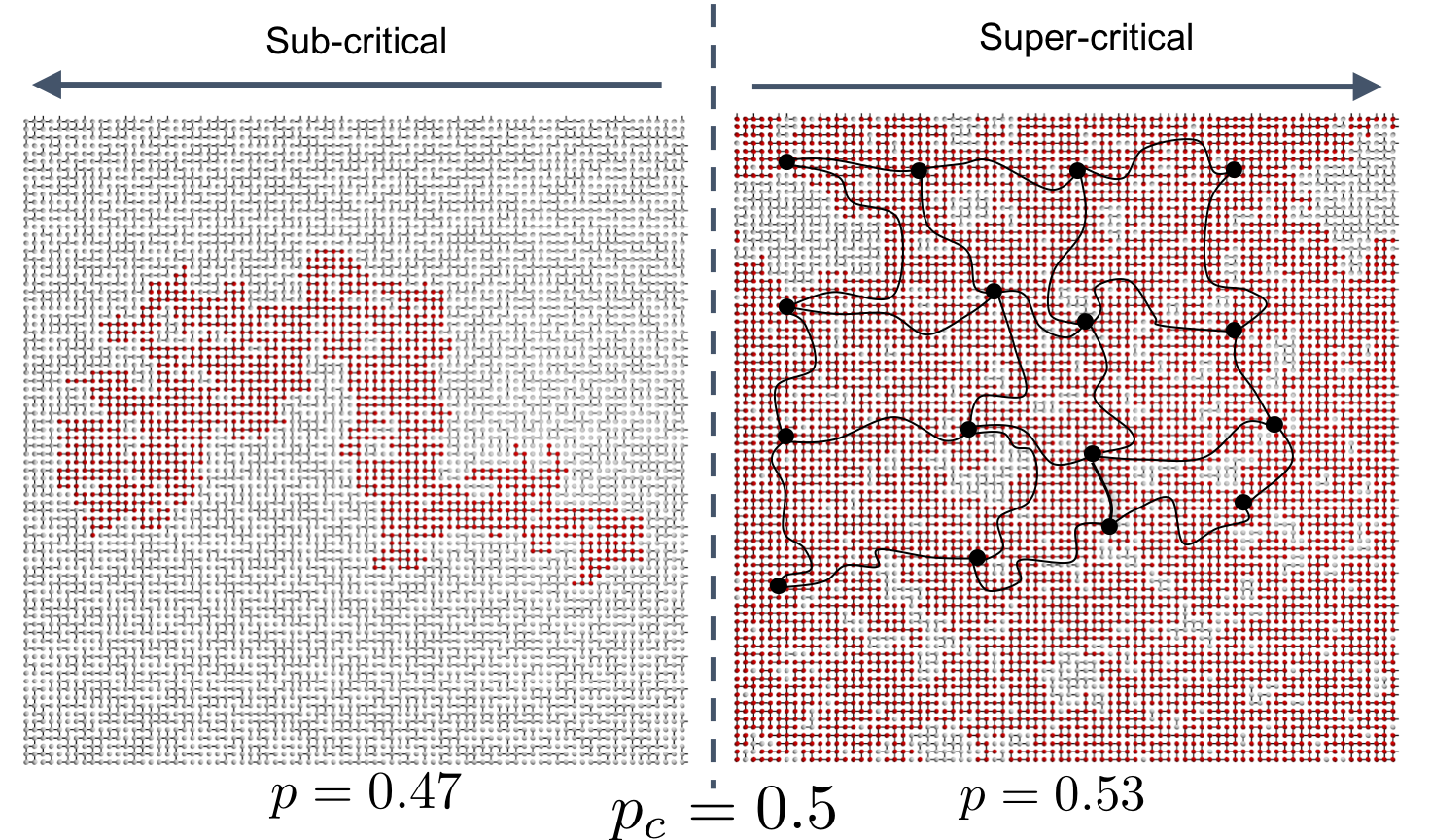}
   \caption{The renormalized grid made by joining the disjoint crossings}\label{fig:crossing}
\end{figure}

Using the rotation invariance of the square lattice under $\frac{\pi}{2}$ rotations, we see that the theorem is true for up-down disjoint crossings of the box with sides of length $n$. This means that above the percolation threshold there exists order $n$ disjoint left-right and order $n$ disjoint up-down crossings of a box $B_{n}$. Also, using translational invariance of the square lattice, we know that this is true for any square box with sides of length $n$. 
Using this result, we now can superimpose a renormalized square grid Fig.~\ref{fig:crossing} using the left-right and up-down disjoint crossings of the square lattice. This ensures that we can have a renormalized square grid of order $\Theta(N)$ sites above the percolation threshold in a bond percolation model. Since all of the intersections of the disjoints crossings on the square grid have at least degree $3$, the renormalized grid ensures the existence of giant $2$-connected and $3$-connected components with sizes of order $\Theta(N)$. However, it does not necessarily guarantee the existence of a giant $4$-connected component since we need order $\Theta(N)$ degree $4$ disjoint crossing intersections in the infinite connected component to guarantee the existence of a giant $4$-connected component. Our numerical results illustrate that the percolation threshold for $4$-connectivity is above the regular threshold for global connectivity.

\begin{figure}[H]
\centering
\begin{subfigure}[b]{0.49\linewidth}
\includegraphics[width=\linewidth]{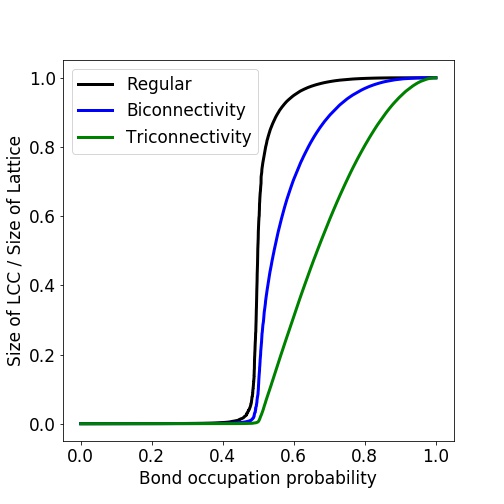}
\caption{}\label{fig:k_conn bond}
\end{subfigure}
\begin{subfigure}[b]{0.49\linewidth}
\includegraphics[width=\linewidth]{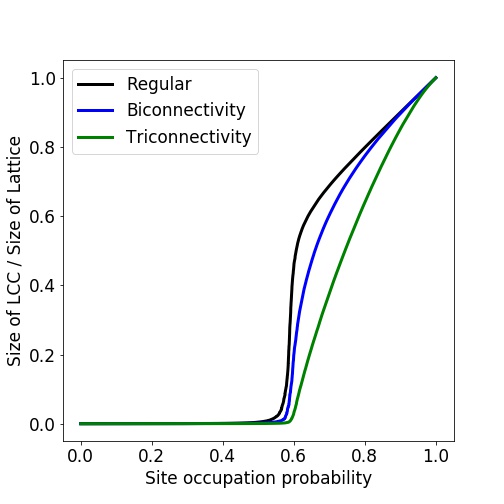}
\caption{}\label{fig:k_conn bond}
\end{subfigure}
\caption{(a) $k$-connectivity for bond percolation on square lattice. The critical threshold is $p_{c}=0.5$. (b) $k$-connectivity for site percolation on square lattice. The critical threshold is $q_{c}=0.5927$}
\label{fig:k-Conn}
\end{figure}

We verified our theoretical results with numerical simulations \cite{hopcroft1973dividing}. Fig.~\ref{fig:k-Conn} shows the percolation threshold for $k$-connectivity for $k=1,2,3$ in a bond percolation on square lattice. As it can be seen from the figure, bond percolation threshold for $k=1$ to $k=3$ is the same and equals $p_{c}=\frac{1}{2}$.

The algorithm for determining the percolation behavior of our model is similar in spirit to the fast percolation algorithm of Newman and Ziff \cite{newman2000efficient} in which we start with an empty graph and randomly occupy bonds one by one until we occupy all the bonds in the square grid. Our simulation results are based on implementations of algorithms developed by Hopcroft et al. \cite{hopcroft1973dividing} and Gutwenger et al. \cite{gutwenger2000linear} which uses depth-first search trees and SPQR trees to find the $2$-components and $3$-components of a given instance of our network.

\subsection{\label{sec:level1}$k$-strong-connectivity and $k$-core}
$k$-strong-connectivity which is less prevalent in the literature and is our strongest robustness measure, demands that every pair of connected nodes in the subset have at least $k$ disjoint paths between them and that nodes on these paths belong to the $k$-strong-component. The $k$-core of a network is obtained by recursively removing the nodes with degree less than $k$ until no such a node exists in the network. $k$-core decomposition has been also applied to many real-world networks (the Internet, the World Wide Web,cellular networks, etc.) \cite{alvarez2005k,wuchty2005evolutionary} and has become an important tool for visualization of complex networks and interpretation of cooperative processes in them. We argue that for a square lattice, $k$-core is the same as $k$-strong-connectivity for all $k$ and is the same as $k$-connectivity for $k=2$.
It is evident that for $k=2$, $k$-strong-connectivity is the same as $k$-connectivity. This is because the nodes that act as conduits in the case of 2-connectivity are parts of cycles in the network and all the nodes in a cycle have $2$ disjoint paths between them and so are $k$-strongly connected. 
We also argue that for $k=3$ on the square lattice, the percolation threshold for $k$-strong connectivity is the same as $k$-core and is equal to 1. 
In order to have a $k$-core, we need all the nodes in the component to have degree $k$ or higher than $k$. The simple square lattice without periodic boundary conditions has four degree $2$ nodes on its corners, in order to have a $3$-core, we need to remove those nodes and continue the process until there are no nodes with degree smaller than three. This results in the deletion of all of the nodes in the square lattice. Therefore, the threshold for $k=3$ for $k$-core and $k$-strong-connectivity is 1 in a square lattice with no periodic boundary conditions.
However, for a square lattice with periodic boundary conditions, the only configuration in which we can have a non-zero $k$-core is a configuration that has a wrap around. The reason is that any configuration that has not wrapped around the torus has a corner that has degree smaller than three which tentatively results in the deletion of all of the nodes in the graph during the $k$-core pruning process. Our numerical results indicate that the $3$-core component and the $3$-strong connected components are exactly the same with the same participating nodes. This is an interesting result since, in general graphs, a $k$-core component does not guarantee being $k$-strong connected. We further numerically investigated the nature of $k$-core transition and the exact point of criticality using finite size scaling. 

\begin{figure}
\centering
\begin{subfigure}[b]{0.49\linewidth}
\includegraphics[width=\linewidth]{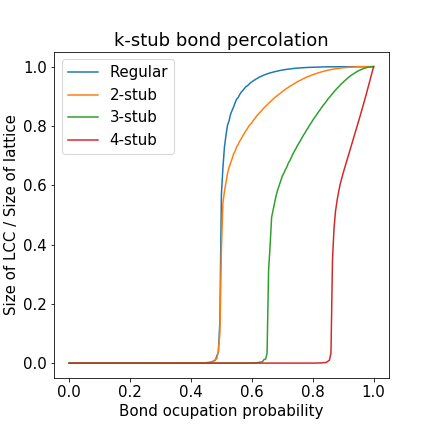}
\caption{}\label{fig:k_conn bond}
\end{subfigure}
\begin{subfigure}[b]{0.49\linewidth}
\includegraphics[width=\linewidth]{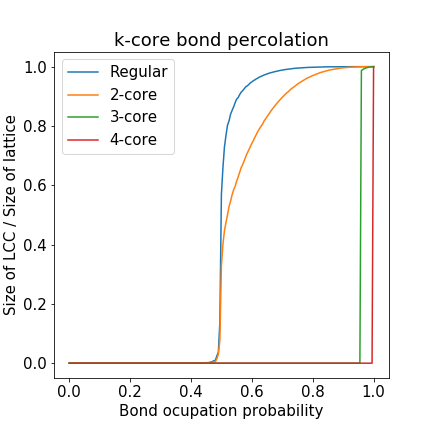}
\caption{}\label{fig:k_conn bond}
\end{subfigure}
\caption{(a) $k$-stub of square lattice for bond percolation. The critical threshold for $4$-stub is smaller than $1$; (b) $k$-core of square lattice. Note that the curve for $2$-core, $2$-connectivity and $2$-strong-connectivity are exactly overlapping}
\label{fig:k}
\end{figure}

\subsection{\label{sec:level1}$k$-stub}
The $k$-stub of a graph is obtained by following the algorithm for finding the $k$-core of the graph only once. We propose an efficient online algorithm similar one described in \cite{newman2000efficient} to find the percolation properties of $k$-stubs of graphs using dynamic updates after the addition of each node. In this algorithm which is based on the union-find data structure, we store extra information on the degrees of nodes. After each node addition, we update the degree of neighboring nodes and check if their degree exceeds $k$. If the degree of any of the neighboring nodes exceeds $k$ after each addition, we add it to the union-find structures that store the $k$-stub components of the graph. This algorithm is not feasible for finding the $k$-core of the graph since the addition of each node could result in non-local changes to the $k$-core components and tracking these nonlocal changes will increase the computational cost of these dynamic updates. 
As can be observed in FIG. 4(a), the percolation threshold for 4-stub is strictly below 1. %We also argue that $q_{c}$, the percolation threshold for 4-stub should be $q_{c}< p_{c}^{1/4}$ where, $p_{c}$ is the site percolation threshold for square lattice. 
\section{Numerical Analysis}
In order to better characterize the nature of robustness measures in our paper, we make use of finite size scaling \cite{landau2014guide,radicchi2010explosive}, a well-known technique developed for numerical analysis of phase transitions. In second order (continuous) phase transitions, every variable $X$ near the critical threshold $p_{c}$ is scale-invariant. This phenomenon appears due to the divergence of correlation length at $p_{c}$. Therefor, $X$ has the following power law form.

\begin{equation}
X \sim \left | p - p_{c} \right |^{\omega }
\end{equation}
where $\omega$ is the critical exponent for variable $X$. On a finite system of size $N$ and length $L$, the variable $X$ has the following scaling form near the threshold
\begin{equation}
X = L^{-\frac{\omega}{\nu}} F\left [ (p-p_{c}) L^{\frac{1}{\nu }} \right ]
\end{equation}
where $\nu$ is the correlation length critical exponent and $F$ a universal function.
At $p=p_{c}$, the scaling function $F$ converges to a constant and variable $X$ follows a simple scaling relation,
\begin{equation}
X \sim \left | L \right |^{-\frac{\omega}{\nu}}.
\end{equation}
Using Monte Carlo simulations of different system sizes at $p=p_{c}$, one can deduce the critical exponent ratio $\frac{\omega}{\nu}$ of the variable using the scaling relation in eq. (3). In this work, following an approach similar to Ref. \cite{radicchi2010explosive}, we adopt the two main variables commonly used to characterize percolation transitions, i.e. the percolation strength $P$ and
the average cluster size $S$. The percolation strength $P$ is defined as the relative size of the largest cluster with respect to the total system size $N$. The scaling relation of $P$ is
\begin{equation}
P = L^{-\frac{\beta}{\nu}} F^{(1)}\left [ (p-p_{c}) L^{\frac{1}{\nu }} \right ]
\end{equation}
Where the critical exponent is $\beta$. $P$ is the order parameter of the transition.
The second variable we use in our numerical simulations is the truncated average cluster size $S$, which is defined as
\begin{equation}
S=\frac{\sum _{s} n_{s} s^{2}}{\sum _{s} n_{s} s}
\end{equation}
In the above equation, $n_{s}$ stands for the number of clusters of size $s$ per node. Since the percolating cluster diverges above $p_{c}$, the sum in eq. (5) runs over all cluster sizes except that of the largest cluster. The scaling relation of $S$ is
\begin{equation}
S = L^{\frac{\gamma}{\nu}} F^{(2)}\left [ (p-p_{c}) L^{\frac{1}{\nu }} \right ]
\end{equation}
Where $\gamma$ is the critical exponent related to average cluster size. In lattice percolation, the exponents $\beta_{L}$, $\nu_{L}$ and $\gamma_{L}$ (where subscript $L$ stands for lattice) are linked by the so-called hyper-scaling relation \cite{stauffer2014introduction},
\begin{equation}
\frac{\gamma}{\nu} + \frac{2\beta}{\nu} = d
\end{equation}
where $d$ is the dimension of the lattice. 

\begin{figure}
\centering
\begin{subfigure}[b]{.49\linewidth}
\includegraphics[width=\linewidth]{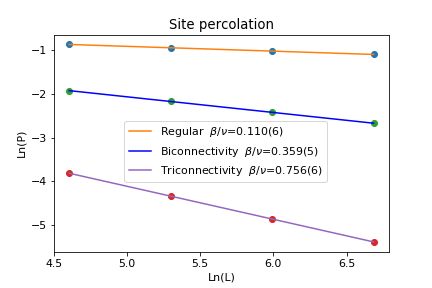}
\caption{}\label{fig:k_conn site beta}
\end{subfigure}
\begin{subfigure}[b]{.49\linewidth}
\includegraphics[width=\linewidth]{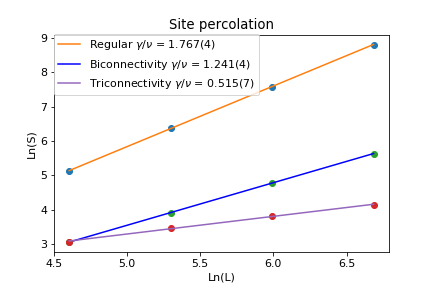}
\caption{}\label{fig:k_conn site gamma}
\end{subfigure}

\begin{subfigure}[b]{.49\linewidth}
\includegraphics[width=\linewidth]{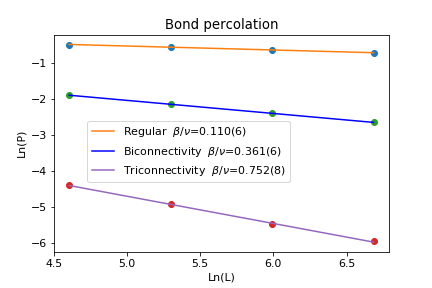}
\caption{}\label{fig:k_conn bond beta}
\end{subfigure}
\begin{subfigure}[b]{.49\linewidth}
\includegraphics[width=\linewidth]{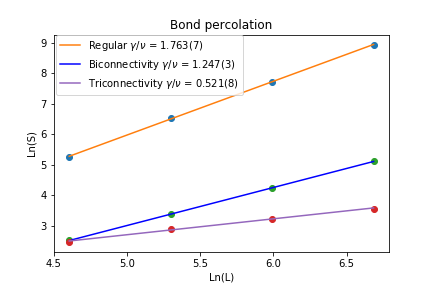}
\caption{}\label{fig:k_conn bond gamma}
\end{subfigure}
\caption{Exponents $\frac{\beta}{\nu}$ and $\frac{\gamma}{\nu}$ for regular percolation, 2-connectivity, and 3-connectivity (a),(b) $p_{c}=0.5927$ and (c),(d) $q_{c}=0.5$. The linear nature of these scaling exponents verifies the fact that we are at the percolation threshold}
\label{fig:exponents}
\end{figure}

\section{RESULTS}

It is known from percolation theory that in regular percolation, the order parameter changes continuously across the transition. This critical behavior is known as a continuous or second order phase transition. On the other hand, first order or continuous phase transitions are characterized by a discontinuity in the order parameter of transitions \cite{ma2018modern}. There is also a class of transitions known as a hybrid transition which are a combination of a first-order discontinuity with a second order transition \cite{azimi2014k}. \\
We examined the critical exponents for all the defined robustness measures. Based on our numerical analysis, the critical behavior of $k$-stub for all $k$ as well as $k$-core and k-strong-connectivity for $k=2$ is similar to regular percolation with similar corresponding exponents. Besides, as shown in the literature \cite{azimi2014k}, our results confirm the fact that $k$-core for $k>2$ is a hybrid phase transition combining a discontinuity and a critical singularity which breaks the usual scenario of ordinary percolation. The critical behavior of $k$-connectivity was found to be different from the percolation universality class. Hence, we report the results corresponding to this robustness measure.
\\
Figure 5. shows the plots for finding critical exponents for bi-connectivity and tri-connectivity, the linear behavior of the curves verifies our theoretical results about the percolation thresholds of these robustness measures on the square lattice. As expected the exponents follow Eq. (7) With a small margin of error which ensures that we indeed deal with a second order phase transition in two dimensions. The values corresponding to $\frac{\beta}{\nu}$ and $\frac{\gamma}{\nu}$ are different from the regular percolation universality class and to the best of our knowledge, they do not belong to any known universality class. Shlifer et al. \cite{shlifer1979large} showed that the correlation length exponent $\nu$ for bi-connectivity is the same as $\nu$ in percolation universality class. Given $\nu=\frac{4}{3}$, we obtain $\beta=0.48(9)$ which is consistent but slightly lower than the previous results, Kirkpatrick \cite{shlifer1979large}, $\beta=0.50(2)$ and Sahimi \cite{sahimi1984scaling}, $\beta=0.542$.
It is also worth noting that the critical behavior of tri-connectivity and bi-connectivity are similar in both bond and site percolation, which is expected since the underlying transition in two dimensions should belong to the same universality class. 
\\
Tables I and II show the percolation thresholds for all our robustness measures. We were not able to find the exact percolation threshold for $4$-connectivity due to its computational complexity; however, our results confirm the fact that that the percolation threshold for this robustness measure is above the regular percolation threshold. 

\begin{table}
\centering
\caption{Bond percolation thresholds}
\label{tab:res}
\bgroup
\def\arraystretch{1.0}
  \begin{tabular}{|c|r|r|r|r|}
    \hline
     \multicolumn{1}{|l|}{} & k-connectivity 				 & k-stub 				 & k-core      &     k-strong-connectivity   \\ \hline
  k=1 & \multicolumn{4}{c|}{0.5}  \\ \hline
    k=2  &   \multicolumn{4}{c|}{0.5}\\ \hline
    k=3     & 0.5 & 0.6603(4) & 0.9692(1) & 0.9692(1)\\ \hline
    k=4  &   & 0.8655(3) & 1  & 1\\ \hline
  \end{tabular}
  \egroup
\end{table}

\begin{table}
\centering
\caption{Site percolation thresholds}
\label{tab:res}
\bgroup
\def\arraystretch{1.0}
  \begin{tabular}{|c|r|r|r|r|}
    \hline
     \multicolumn{1}{|l|}{} & k-connectivity 				 & k-stub 				 & k-core      &     k-strong-connectivity   \\ \hline
  k=1 & \multicolumn{4}{c|}{0.5927}  \\ \hline
    k=2 & \multicolumn{4}{c|}{0.5927} \\ \hline
    k=3     & 0.5927 &0.7356(3) &0.9747(1) & 0.9747(1)\\ \hline
    k=4  &   & 0.8846(4) &  1 & 1\\ \hline
  \end{tabular}
  \egroup
\end{table}

\section{Conclusion}
We numerically evaluated various percolation phenomenon for several robustness measures for the square grid. Our results show that $k$-stub for all $k$  and $k$-strong-connectivity and $k$-core for $k=2$ belong to the percolation universality class in two dimensions. Critical exponents for $k$-connectivity is shown to belong to different universality classes for different $k$ and the corresponding exponents for $k=2,3$ are calculated. In addition, we report percolation thresholds for all the robustness measures and show that on the square lattice, the percolation threshold for $k$-connectivity for $k=2,3$, and $k$-core, $k$-stub,   and $k$-strong-connectivity for $k=2$ are equal to the ordinary percolation threshold. In ongoing work, we are translating these connectivity-based network robustness measures to understand how to design and control a software-defined wireless network to realize distributed analytics that is robust to network dynamics.

\section{Acknowledgements}
This research was sponsored in part by the U.S. Army Research Laboratory and the U.K. Ministry of Defense under Agreement Number W911NF-16-3-0001. The views and conclusions contained in this document are those of the authors and should not be interpreted as representing the official policies, either expressed or implied, of the U.S. Army Research Laboratory, the U.S. Government, the U.K. Ministry of Defense or the U.K. Government. The U.S. and U.K. Governments are authorized to reproduce and distribute reprints for Government purposes notwithstanding any copyright notation hereon.

\renewcommand\refname{Bibliography}
\newcommand{\newblock}{}
\bibliographystyle{plainnat} % or try abbrvnat or unsrtnat
\bibliography{BIB} % refers to example.bib

\end{document}